# Probabilistic Models for Daily Peak Loads at Distribution Feeders

Hossein Sangrody, *Student Member, IEEE*
Department of Electrical and Computer Engineering
Binghamton University
Binghamton, NY 13902, USA
habdoll1@binghamton.edu

Ning Zhou, *Senior Member, IEEE*
Department of Electrical and Computer Engineering
Binghamton University
Binghamton, NY 13902, USA
ningzhou@binghamton.edu

Xingye Qiao
Department of Mathematical Sciences
Binghamton University
Binghamton, NY 13902, USA
qiao@math.binghamton.edu

***Abstract*— Load forecasting at distribution networks is more challenging than load forecasting at transmission networks because its load pattern is more stochastic and unpredictable. To plan sufficient resources and estimate DER hosting capacity, it is invaluable for a distribution network planner to get the probabilistic distribution of daily peak-load under a feeder over long term. In this paper, we model the probabilistic distribution functions of daily peak-load under a feeder using power law distributions, which is tested by improved Kolmogorov–Smirnov test enhanced by the Monte Carlo simulation approach. In addition, the uncertainty of the modeling is quantified using the bootstrap method. The methodology of parameter estimation of the probabilistic model and the hypothesis test is elaborated in detail. In the case studies, it is shown using measurement data sets that the daily peak-loads under several feeders follow the power law distribution by applying the proposed testing methods.**

***Index Terms*—load forecasting, daily peak load probabilistic model, power law distribution, empirical load modeling, probabilistic forecasting**

## I. Introduction

Load forecasting (LF) plays a critical role in power system control, operation, planning, and marketing. Nowadays distributed energy resources (DERs) are penetrating into the grid at an accelerating speed, which present unprecedented challenges to power grid operators because they introduce more and more indispatchability, variability, uncertainty, and bi-directional power flow to power grids [1]. As such, it is vital to have a robust LF tool that can provide a view of ongoing and future conditions of local distribution grids.

Conventionally, LF research is focused on providing point estimation of target load [2]. However, there are some decision making applications such as probabilistic load flow [3], electricity market [4], unit commitment [5], and reliability planning [6] which rely on probabilistic load forecasting (PLF). In addition, the uncertainty of forecasting model inputs (weather and economic indicators) [7, 8], and the increasing penetration of DERs which imposes high variability into power distribution system cause more attention to PLF.

In [7] density forecast for weekly and yearly peak demands is represented. A comprehensive tutorial review on PLF is discussed in [9]. With the emergence of smart meters providing high geo-resolution data, researchers tend to take benefits of such data in their LF of local loads. In [10] and [11], short-term load forecasting (STLF) with smart meters is applied while the accuracy of LF is improved by customer clustering. In [8], PLF using advanced metering infrastructure (AMI) data is presented with the aid of economic and weather data scenarios.

Load forecasting is even more challenging for small scale networks because the variability of the load pattern increases when networks' scale decreases. Accordingly, it is more challenging to obtain accurate LF at distribution level than at transmission level whose load patterns are smoother. In addition, the load pattern is becoming more stochastic with the presence of DERs, especially with renewable generations such as rooftop solar PVs.

To plan sufficient resources and estimate DER hosting capacity, it is invaluable for a distribution network planner to have the probabilistic distribution of daily peak-load under a feeder over long term. To address the needs, we use the Kolmogorov–Smirnov test which is enhanced by Monte Carlo simulations to verify potential probabilistic distribution functions of daily peak-load under a feeder. The methodology of deriving optimal parameters of the probabilistic model and performing hypothesis test is elaborated. It is shown that the measured loads under several tested feeders follow the power law distribution. The proposed probabilistic model can be implemented to derive a PLF model for long-term load forecasting for distribution networks and feeders with AMI data.

The rest of the paper is presented as follows. In Section II, a description of the problem is represented. Section III elaborates the methodology of modeling empirical data based on power law distribution. Simulation results on several case studies and the conclusion are represented in Section IV and V, respectively.

## II. PROBLEM DISCRIPTION

The complementary cumulative distribution function (CCDF), or the survival function, of the distribution of peak load at $x$ is the probability that the (random) peak load is greater than or equal to $x$, $S(x) = P(X \geq x)$. As shown in Table I, for an observed data set, the empirical version of the CCDF at $x$ is the sum of the frequencies of those observed peak loads that are greater than or equal to $x$ over the total sample size $N$. Such a peak load may be considered for hourly, daily, weekly, monthly, yearly, or any other arbitrary term.

According to this representation of empirical data, if there is a distribution model whose CCDF can completely or partially mimic the empirical CCDF of the peak load, a forecaster may be able to predict the probability of having a peak load excessing a threshold value. In the next section, we present the methodology to model the CCDF of peak load with a distribution called the power law distribution.

TABLE I. PEAK LOAD CCDF TABLE

| PEAK LOAD RANKING | PEAK LOAD FREQUENCY | CCDF |
|---|---|---|
| **Largest Peak Load** | $f_1$ | $\frac{f_1}{N}$ |
| **Second Largest Peak Load** | $f_2$ | $\frac{(f_1+f_2)}{N}$ |
| ⋮ | ⋮ | ⋮ |
| **Second Smallest Peak Load** | $f_{N-1}$ | $\frac{(f_1+f_2+\cdots+f_{N-1})}{N}$ |
| **Smallest Peak Load** | $f_N$ | 1 |

## III. PROPOSED METHODOLOGY

As mentioned in the last section, if the observed peak load can be modeled with the CCDF of a distribution, a probabilistic peak load model is achievable. Such a probabilistic model may present the empirical CCDF completely, or partially such as at the tail of the CCDF.

For any term, there might be different distribution representing the CCDF of peak load pattern. In this paper, it will be shown that the CCDF of daily peak load can be modeled partially with power law distribution. Thus, from this point on, peak load refers to daily peak load.

There are a lot of natural or man-made events which follow power law distribution. For example, both the peak gamma-ray intensity of solar flares during 1980 to 1989 and the number of customers affected by energy outage in the united states during 1989 to 2002 can be well modeled by the power law distribution [12]. An example of the power law distribution with application in forecasting is the earthquake magnitude distribution modeling in [13].

A random variable follows the power law distribution if its probability density function (PDF) can be described by (1),

$$p(x) \propto x^{-\alpha} \qquad (1)$$

where $\alpha$ is a scaling parameter [12]. The function $p(x)$ in (1) represents the PDF of the power law distribution at any possible value $x$. However, in many real cases, only the right tail of distribution, that is, those greater than or equal to some value $x_{min}$, follows the power law distribution closely. In this case, we let $W = S(x_{min}) = P(X \geq x_{min})$ be the CCDF at $x_{min}$. Then (1) can be modified to (2),

$$p(x) = W\frac{\alpha-1}{x_{min}}\left(\frac{x}{x_{min}}\right)^{-\alpha} \quad \text{for } x \geq x_{min} \qquad (2)$$

The CCDF of the power law distribution function is then (3).

$$S(x) = W\left(\frac{x}{x_{min}}\right)^{-\alpha+1} \quad \text{for } x \geq x_{min} \qquad (3)$$

Observe that the CCDF in (3) is defined for $x \geq x_{min}$ only and it relies on $W = S(x_{min})$, which can be consistently estimated by the empirical CCDF based on the data. Note that our focus is on the right tail of the distribution, that is, the chance that the peak load gets too big; therefore, it is not our primary concern to model the entire CCDF of the peak load distribution. Although (3) is not the CCDF of power law distribution, it is referred to as the CCDF of power law distribution in this paper for the simplicity of discussion.

To check whether the data fits the power law distribution, it is necessary to first estimate the parameters of power law distribution, $x_{min}$ and $\alpha$.

### A. *Estimating Power Law Distribution Parameters*

We briefly review how to estimate $x_{min}$ and $\alpha$ here. Each observed value peak load in the data set is considered as potential values for $x_{min}$ one by one and the observed values less than $x_{min}$ is truncated from the data set. Then, using the remaining data of $x \geq x_{min}$, two CCDFs, i.e., the empirical CCDF and the theoretical CCDF are obtained. The empirical CCDF is obtained from the frequency table as it was described in Section II while the theoretical CCDF for power law distribution is described in (3) (after setting $W$ =1) in which $\alpha$ is estimated using the maximum likelihood principal based on the truncated data. The differences between the theoretical and empirical CCDFs are assessed using Kolmogorov–Smirnov (KS) test statistic as represented in (4). Note that the KS test statistic finds the largest distance $D$ between the theoretical and empirical CCDFs.

$$D = \max_{x \geq x_{min}} \left(\left|CCDF_{theortical} - CCDF_{emperical}\right|\right) \qquad (4)$$

The results of KS test for all $x_{min}$ is a vector. $x_{min}$ leads to the smallest KS test statistic is our best guess for this parameter. In addition, the other parameter $\alpha$ can be obtained using maximum likelihood estimation (MLE). See more details in [12].

### B. *Goodness of Fit for the Probabilistic Model*

As explained, the KS test statistic and the maximum likelihood estimation (MLE) can be used to derive the parameters of a power law distribution, which seems to match the data set very well. However, fitting a power law distribution to a data set does not mean that the power law distribution is indeed the distribution for the observed data. It is necessary to conduct a goodness-of-fit test to check whether the observed data $x$ follow the power law distribution. The null hypothesis of the test is set as "the observed data follow the power law distribution" [14].

Any off-the-shelf goodness-of-fit test (such as the K-S test) can assess the discrepancy between the CCDFs of the empirical data and the power law distribution with some parameter values, and generate a $p$-value. Assume that the significance level of the test is set 10%. Then, if the $p$-value is less than 10%, the null hypothesis is rejected (i.e., the data set does not follow the power law distribution). It indicates that when the null hypothesis is true, there is less than 10% chance that we may observe a data set whose test statistic is more extreme than the currently observed one. However, these tests state the null hypothesis conclusion by testing empirical data and the potential distribution whose parameters are estimated from the same empirical data. Therefore, the test may be invalid since it may favor the null hypothesis. Thus, it is necessary to improve the test so that it is valid. One such approach is by Monte Carlo.

To obtain a valid $p$-value using Monte Carlo, a large number (say 10,000) of synthetic random samples with the same size as the original data is generated from the power law distribution with the same estimated parameter values. Then, for each simulated sample, a power law distribution is fitted and its corresponding KS statistic under the null hypothesis that the data come from power law distribution with the newly estimated parameter values is obtained. The proportion of the KS statistics for the simulated data that are less than or equal to the KS statistic for the original data is the improved $p$-value. This $p$-value is valid because it takes into account the fact that the parameter values are estimated using the empirical data.

Note that the empirical data may follow other distribution such as exponential, gamma, lognormal, etc. Tests of other distributions can also be performed using the same method as above and are also carried out in this paper.

In addition, the uncertainty of the fitted distribution model is quantified using bootstrapping. Bootstrapping is a nonparametric approach to estimate the statistical parameters from a sample through resampling with replacement. By bootstrapping the empirical data, the variation of fitted model is obtained in a confidence interval (CI) [15]. CIs of 95% are considered in this study.

## IV. Simulation Results

In power system, the peak load is usually considered in hourly, daily, weekly, monthly, and yearly time frame, although it may be considered in any arbitrary time frame as well. The power law distribution model is tested for hourly, daily, weekly, and monthly on 4 local private networks. The results of simulations indicate that among the aforementioned time frames, daily peak load follows power law distribution.

The methodology represented in Section II and III are elaborated by simulation on one of the local case study networks using MATLAB® and the results of simulation on other case studies are also shown briefly with table and figures. All the local networks supply 2000 to 3500 customers including residential, commercial, and industrial classes with smart meters for two years. To protect the confidentiality, the load levels are scaled and time plots are not shown in this paper. The daily peak load data are sorted and listed in Table II, which corresponds to the empirical CCDF shown in Table I.

Table II. Empirical CCDF of Daily Peak Load

| Load Ranking | Load Value | CCDF |
|---|---|---|
| **Largest Peak Load** | 3721 | 0.0014 |
| **Second Largest Peak Load** | 3668 | 0.0034 |
| . . . | | . . . |
| **Second Smallest Peak Load** | 1780 | 0.9987 |
| **Smallest Peak Load** | 1577 | 1 |

As shown in Table II, the CCDF in the third column represents the empirical CCDF of the daily peak load for the case study. The CCDF of 0.0014 corresponding to the largest load 3721 indicates that the probability of having a peak load greater than or equal to 3721 is 0.14% for a day. To have a probabilistic model for the continuous range of daily peak load and having CIs of probability forecasting, the power law distribution is tested using the current empirical data set.

According to the method described in Section III, the parameters of power law distribution, i.e. $x_{min}$ and $\alpha$, whose CCDF fits the empirical CCDF are obtained using the KS test and MLE, respectively. The number of bootstrapping for CIs and Monte Carlo repetition are 2500 [12]. Fig. 1 illustrates the results of KS test. As shown in this figure, the KS statistic reached its minimum distance at $x_{min}$=3085.

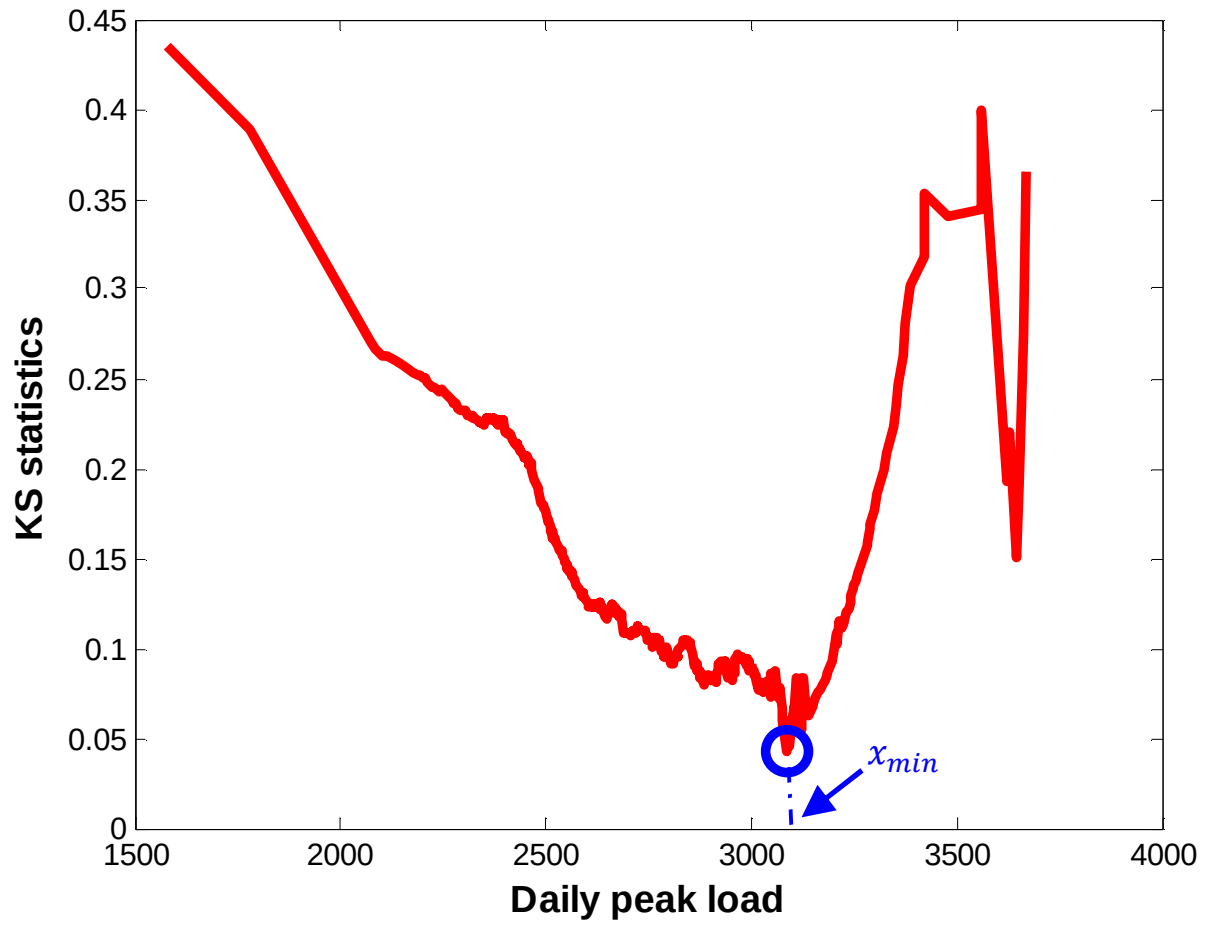

Fig. 1. KS statistics results for Monte Carlo repetitions

The estimated parameters, CIs, and the $p$-value are summarized in Table III. As shown in this table, $p$-value is greater than 0.1, which means that the null hypothesis cannot be rejected. In addition, the empirical data is tested on other similar distributions like lognormal, exponential, gamma, etc. and their $p$-values reject the null hypothesis.

The power law distribution model for the CCDF of the daily peak load are shown in Fig. 2. In this figure, the blue dot represented the empirical CCDF and the red dashed line shows the theoretical CCDF of the power law distribution estimated from the empirical data.

TABLE III. POWER LAW DISTRIBUTION MODEL FOR FIRST CASE STUDY

| PARAMETERS OF POWER LAW DISTRIBUTON | PARAMETER VALUE | CI, 95% | P-VALUE |
|---|---|---|---|
| $x_{min}$ | 3085 | $2786 \leq x_{min} \leq 3173$ | 0.758 |
| $\alpha$ | 22.28 | $14.4 \leq \alpha \leq 29$ | |

Fig. 3 shows a zoom-in plot of the red-frame area in Fig. 2 whose empirical data is modeled based on the power law distribution. In both figures, the CIs of 95% are shown with green dashed lines and shadows.

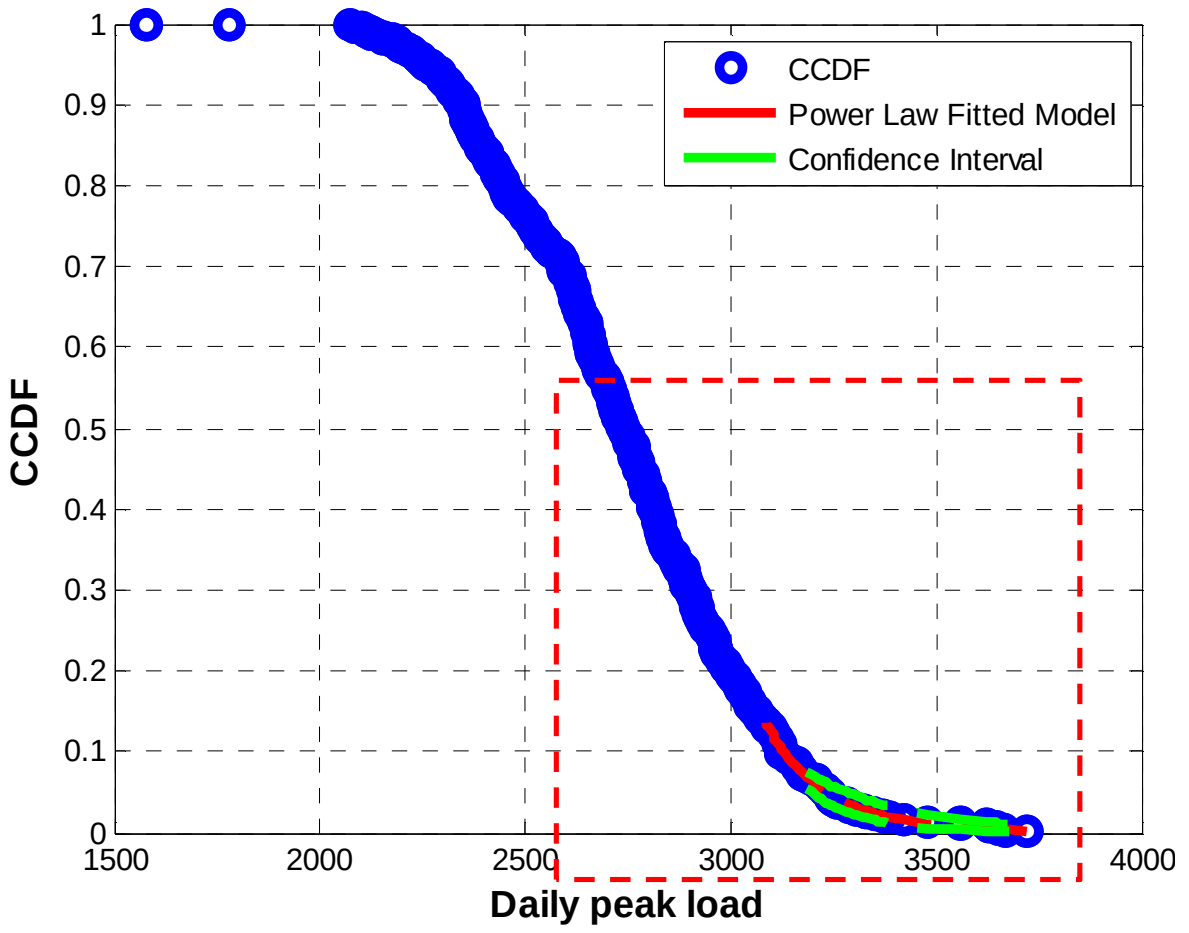

Fig. 2. Power law distribution model for empirical data of case study #1

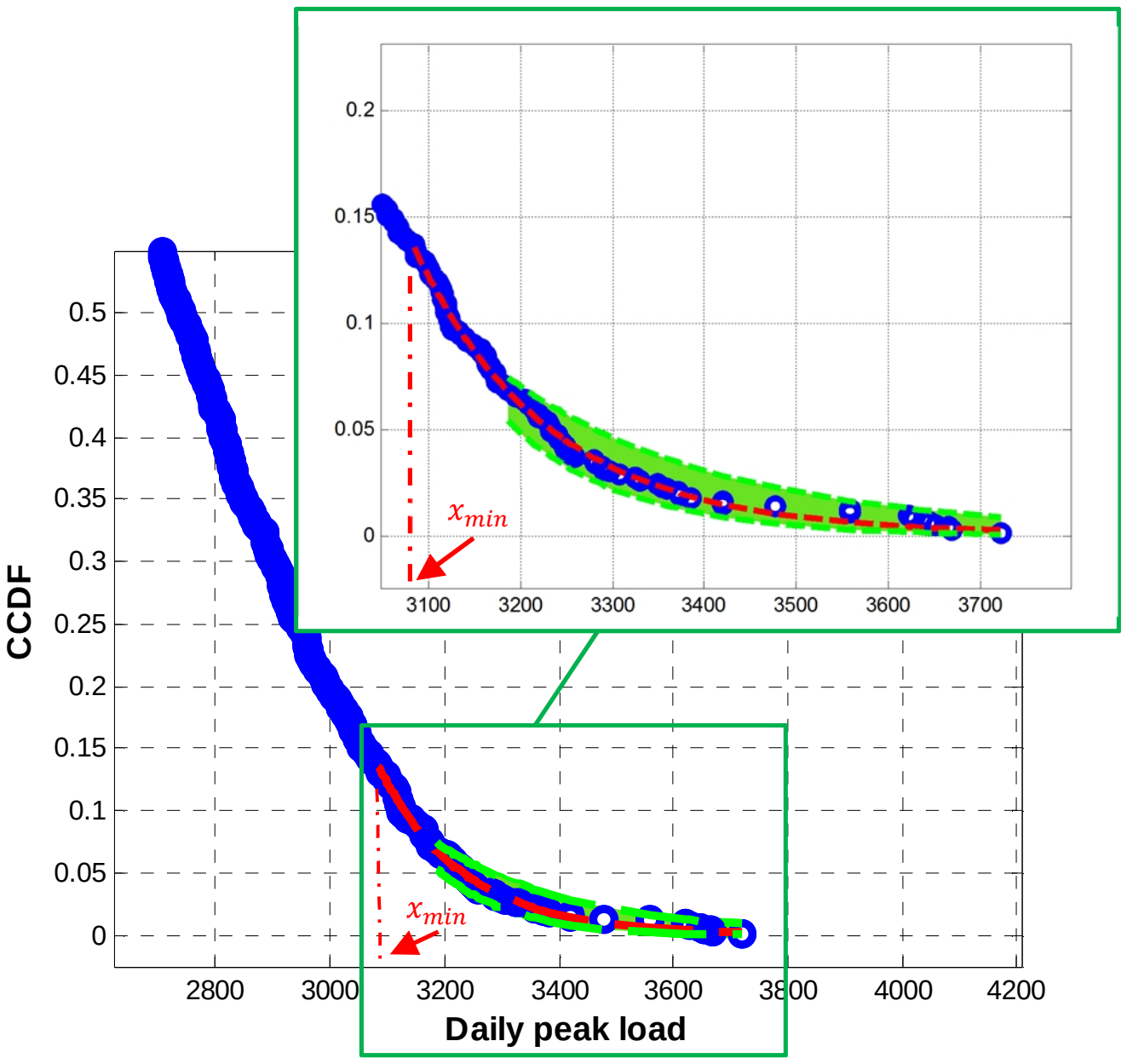

Fig. 3. Power law model with confidence intervals for case study #1

The empirical data in Fig. 3. is modeled with the power law distribution starting at 3085 Wh. In other words, the daily peak loads, whose magnitudes are greater than 3085 Wh, begin to follow the power law distribution model. Note that the peak loads less than 3085 Wh is not modelled in this case. However, in the most of LF applications, high peak-load demand is more important than low peak-load demand.

As seen in Fig. 3, the probability of any daily peak load great than or equal to $x_{min}$ is obtained using (3). For example, considering the parameters of the current power law model in Table III, the probability of a daily peak load being greater than or equal to 3400 Wh is 0.0172 ($W$ is 0.1356 for this case) and the 95% confidence interval ranges from 0.0116 to 0.0298. Chance of peak load at other value can be calculated with the same procedure. Fig. 4. depicts another view of power law distribution modeling on a logarithmic scale.

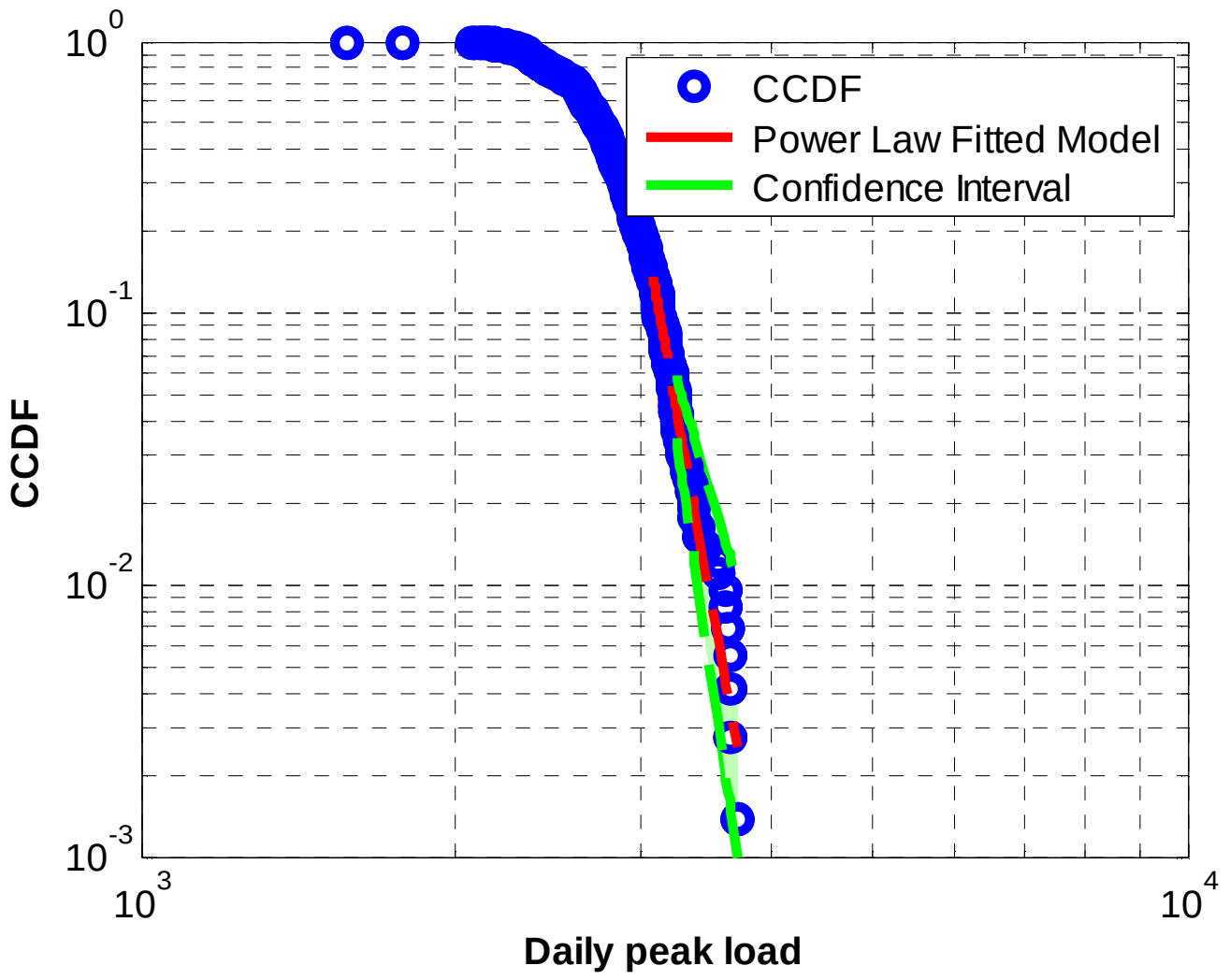

Fig. 4. Logarithmic illustration of power law modeling on empirical data

The methodology of power law modeling is also applied on 3 other case studies. As it was mentioned earlier, the empirical data is collected by aggregating AMI data of the local network with 2000 to 3500 customers. The result of the simulations is shown in Table IV. As shown in the table, the *p*-values of the all case studies are greater than 0.1.

TABLE IV. POWER LAW DISTRIBUTION MODEL FOR OTHER CASE STUDIES

| NETWORK | PARAMETERS | VALUE | PARAMETER CI 95% | P-VALUE |
|---|---|---|---|---|
| Case Study #2 | $x_{min}$ | 2204 | $1481 \leq x_{min} \leq 2351$ | 0.41 |
| | $\alpha$ | 21.88 | $6.046 \leq \alpha \leq 27.068$ | |
| Case Study #3 | $x_{min}$ | 4490 | $4350 \leq x_{min} \leq 4735$ | 0.51 |
| | $\alpha$ | 19.8 | $16.58 \leq \alpha \leq 26.64$ | |
| Case Study #4 | $x_{min}$ | 4158 | $3755 \leq x_{min} \leq 4299$ | 0.45 |
| | $\alpha$ | 24.17 | $14.58 \leq \alpha \leq 30.44$ | |

Figs. 5-7 show a section of empirical data for the three case studies, in which the power law distribution is used to model the load data under different feeders.

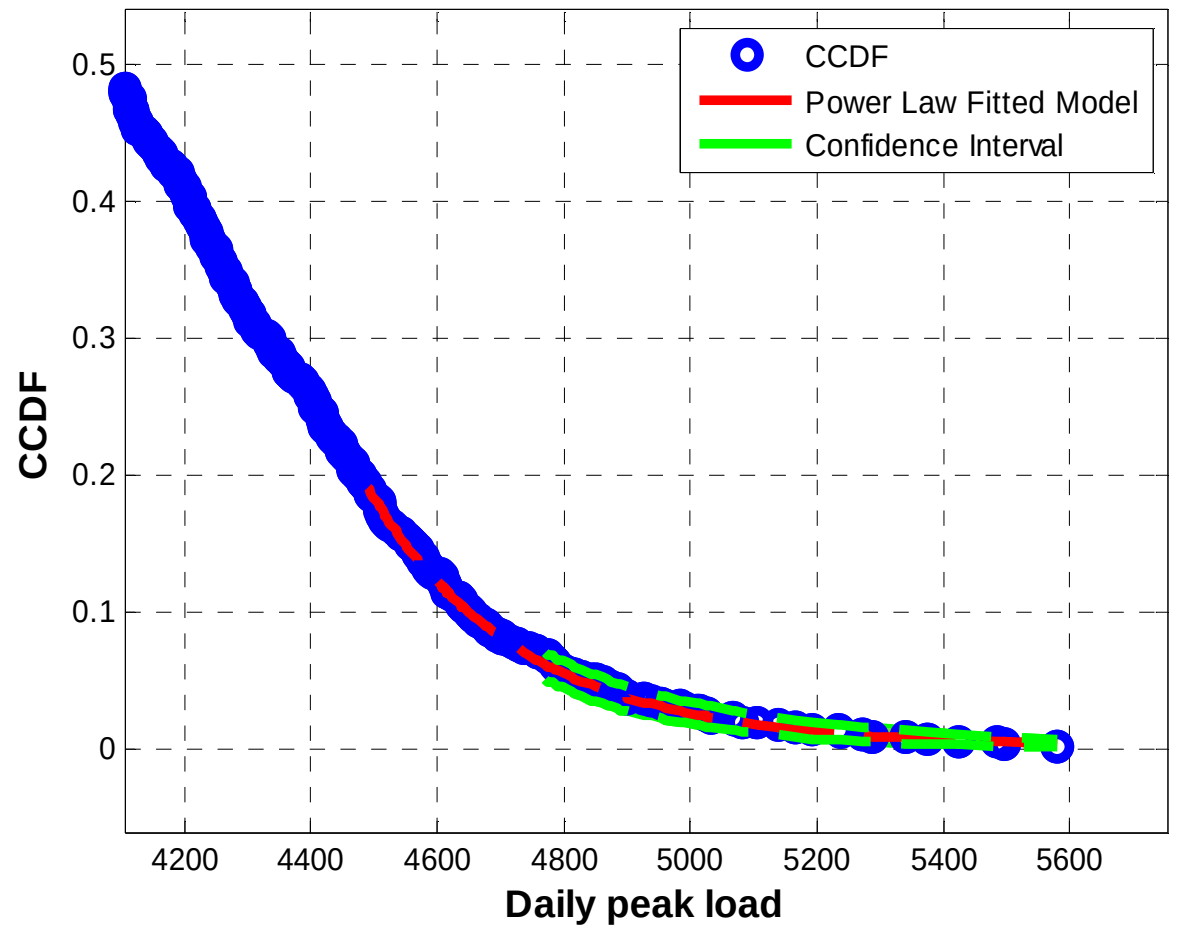

Fig. 5. Power law distribution fit on daily peak load of case study #2

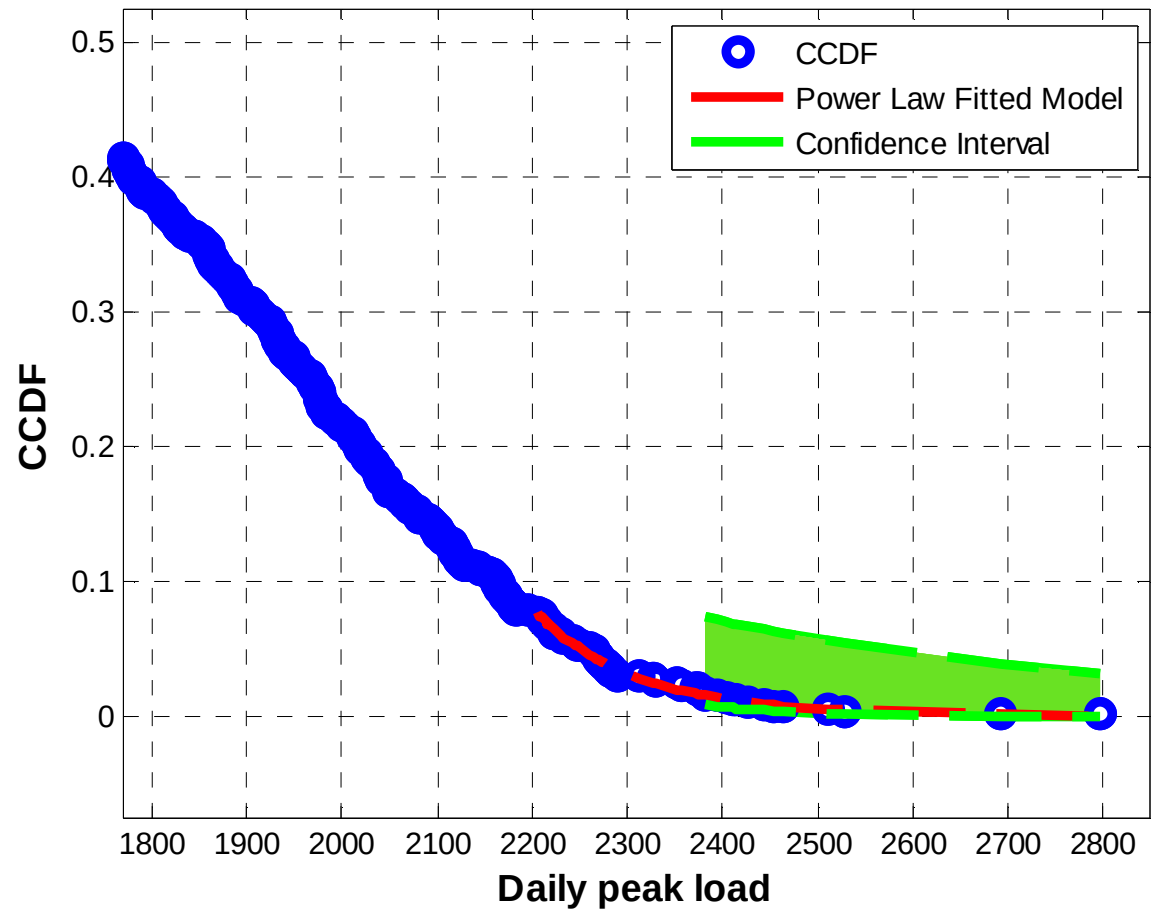

Fig. 6. Power law distribution fit on daily peak load of case study #3

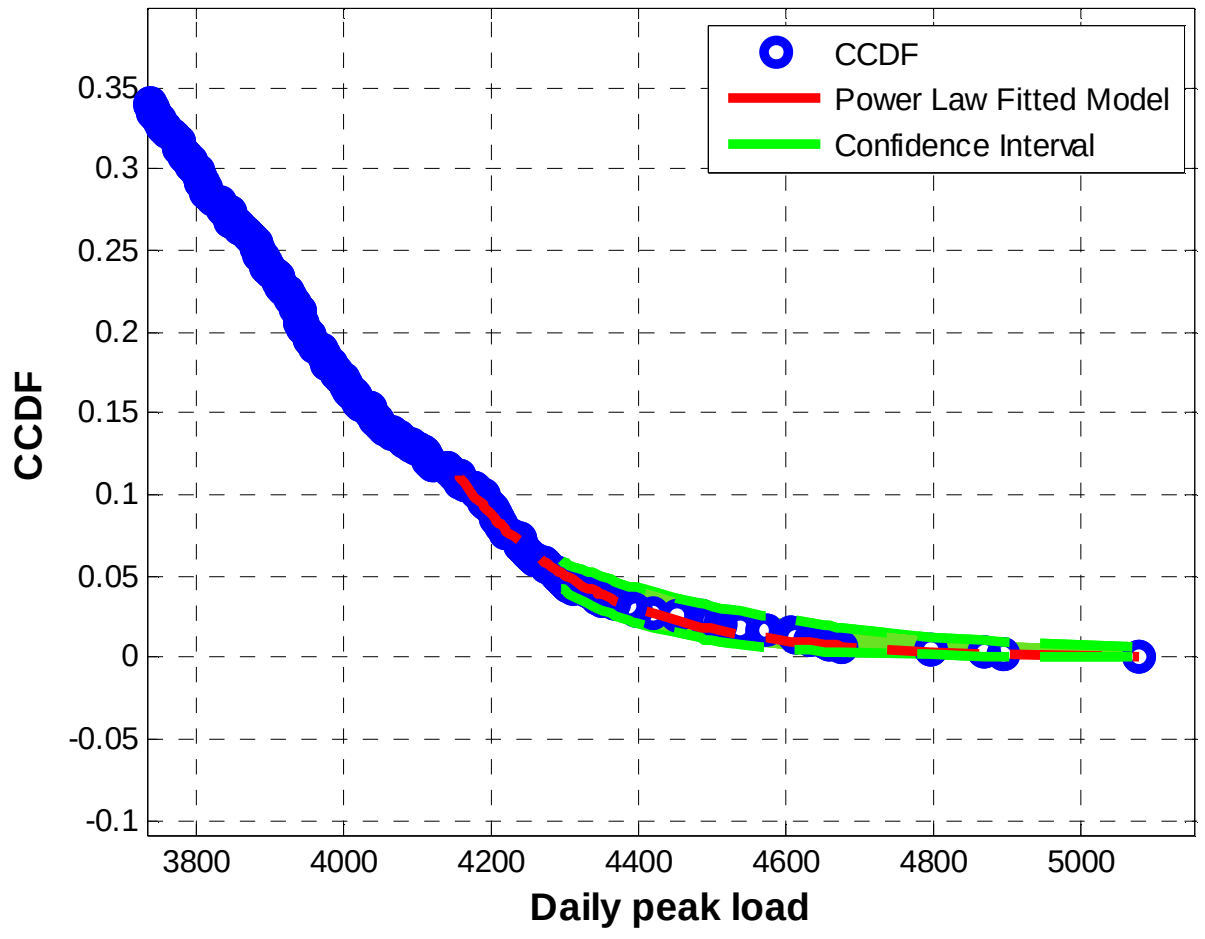

Fig. 7. Power law distribution fit on daily peak load of case study #4

Note that the methodology of power law distribution modeling is based on historical data of a two-year window. Such historical data should be updated in the window frame to cover any rising or falling trend. The probabilistic daily peak model represented in this paper can be implemented to derive a PLF model for long-term load forecasting for distribution networks and feeders with AMI data or any probabilistic applications

## V. Conclusion

This paper proposes the Monte-Carlo method and KS test to verify the probabilistic distributions of the daily peak load. In addition, the bootstrap method is applied to obtain the confidence intervals of the CCDF. Applying the proposed methods, it is shown through the case studies that daily peak-load under several feeders follows the power law distribution. The proposed method can be used to address the challenges of long-term small-area load forecasting.